\documentstyle[sprocl]{article}

    \def\be{\begin{equation}}
    \def\ee{\end{equation}}
    \def\bea{\begin{eqnarray}}
    \def\eea{\end{eqnarray}}

    \title{The Spirit and the Letter of Copenhagen: \\A Response to Andrei Khrennikov}
    \author{Arkady Plotnitsky}

\address{Theory and Cultural Studies Program \\Department of English
\\Purdue University \\West Lafayette, IN 47907 \\e-mail:
aplotnit@sla.purdue.edu }
    
\begin{document}
    \maketitle

    \abstracts{The present communication addresses (by way of a
    response to Andrei Khrennikov's recent argument), the epistemology
    of quantum mechanics and Bohr's interpretation of quantum
    mechanics as complementarity.}

\section{Introduction}
    
In his posting to the ArXiv (quant-ph/0202107 v1), Andrei Khrennikov
offers what he calls the ``V\"{a}xj\"{o} Interpretation of Quantum
Mechanics'' (alluding perhaps to David Mermin's ``Ithaca
Interpretation'') along with a critical commentary on ``the Copenhagen
Interpretation'' of quantum mechanics, and, in Heisenberg's famous
phrase, the spirit of Copenhagen, and specifically Niels Bohr's
views.\footnote{ Andrei Khrennikov, ``V\"{a}xj\"{o} Interpretation of
Quantum Mechanics,'' arXiv: quant-ph/0202107 v1, 19 Feb 2002; On
Mermin's ``Ithaca Interpretation,'' see N. David Mermin, ``What Is
Quantum Mechanics Trying To Tell Us?''  {\it American Journal of
Physics} 66 (9), 1998: 753-67.} As his commentary refers, at arguably
crucial points, to my own discussions of these subjects, I would like
to address some of his points here.  I also hope my remarks will help
to clarify at least some of the aspects of Bohr's interpretation of
quantum mechanics, known as complementarity, about which, {\it the
spirit and the letter of which}, there are still so many
misconceptions and so much confusion.  One might take advantage of the
pun on the very word ``letter'' and see Bohr's writings on
complementarity as a letter or several letters, which were written to
Einstein in particular, but also to all of us, and which are still
awaiting their recipients at the post office of the history of quantum
theory.  In this case (or perhaps in any case), however, having
received and even having read these letters by no means guarantees
that they have reached their destination, including in the case of the
present author.  It is not entirely clear, and it did not appear to
have been to Bohr, to what degree these letters reached their
destination in Einstein's case.  But they have reached their destiny
insofar as the Bohr-Einstein debate has fundamentally shaped the
history of quantum theory.

It would be difficult for me to assess the merits of the V\"{a}xj\"{o}
interpretation itself, since, as an interpretation, it does not appear
to me to be sufficiently explicated in Khrennikov's communication as
an interpretation of quantum mechanics as a physical theory. 
Accordingly, I shall not address it here.  One certainly cannot object
to pursuing this type of program; and I find, in particular, the {\it 
p}-adic approach to quantum theory intriguing, at least from a
mathematical standpoint, even if only because the mathematics of {\it
p}-adic numbers is very beautiful and quantum considerations may well
contribute to this mathematics.  On the other hand, I found
Khrennikov's argument concerning complementarity puzzling.  The nature
of my puzzlement stems primarily from my view of complementarity as an
interpretation of quantum mechanics, and I shall restate this view
first, in the first part, ``The Spirit of Copenhagen,'' of this paper. 
My position is, by and large, similar to that of Bohr or Heisenberg,
although one might argue that they, especially Heisenberg, sometimes
made stronger claims for complementarity as an interpretation of
quantum mechanics.  There are also those who see complementarity or
other interpretations of quantum mechanics as actual descriptions of
nature at the quantum level rather than as specific interpretations of
a particular theory or experimental data it deals with and only in
this, more limited, way telling us something about nature or about
ourselves in our interaction with nature through our experimental
technology.  What these interpretations (there is no quantum mechanics
apart from them) tell us is of course the question.  Then, I shall
address certain key aspects of complementarity that do not appear to
me to be sufficiently taken into account by Khrennikov's paper.  In
the second part, ``The Letter of Copenhagen,'' I shall comment on
several specific points of the paper, in particular those dealing with
Bohr's actual statements.\footnote{ I permit myself to refer to my
previous works and further references in them, specifically ``Reading
Bohr: Complementarity, Epistemology, Entanglement, and Decoherence,''
{\it Decoherence and its Implications for Quantum Computation and
Information Transfer}, T. Gonis and P.E.A. Turchi (Eds.), IOS Press,
2001; Chapter 2, ``Quantum Mechanics, Complementarity, and Nonclassical
Thought,'' of my book, {\it The Knowable and the Unknowable: Modern Science,
Nonclassical Thought, and the ``Two Cultures,''} University of Michigan
Press, 2002, pp.  29-108; and ``Quantum Atomicity and Quantum
Information: Bohr, Heisenberg, and Quantum Mechanics as an Information
Theory,'' forthcoming in the proceeding of the conference, {\it Quantum
Theory: Reconsideration of Foundations}, V\"{a}xj\"{o}, Sweden, 2001.  I shall,
however, try to make my argument here sufficiently self-contained to
be read without consulting these works.}

\section{The Spirit of Copenhagen}

I see complementarity is {\it an} interpretation, a particular
interpretation, of quantum mechanics, but only one among other
possible interpretations, of which there are now many (the assessment
of their effectiveness is of course a different matter).\footnote{ It
is worth further specifying that by quantum mechanics itself I only
mean the theory covered by the standard mathematical formalism in
whatever version, beginning with HeisenbergÕs matrix formalism, rather
than, say, any of the Bohmian hidden-variables theories.  Technically,
there are further epistemological and even physical, and hence also
interpretive, complexities defined by particular versions of the
mathematical formalism of quantum mechanics, such as that of
Heisenberg vs.  that of Schr\"{o}dinger, but I shall put these
complexities aside here.  In any event, the formalism in question is
seen as enabling the predictions of the data in question, such as the
results of the double-slit experiments, the EPR correlations, and so
forth.} Indeed, there is more than one version of complementarity,
even in Bohr's own work, let alone if one considers the work of others
who appeal to complementarity, several founders of quantum mechanics
among them.  These interpretations are sometimes assembled under the
rubric of the Copenhagen or, as it also called, orthodox
interpretation, and often uncritically seen as a single
interpretation.  Instead, the phrase ``the Copenhagen Interpretation''
may be best seen as referring to a cluster of related interpretations
of quantum mechanics, which do share certain physical and
epistemological features, but which also, sometimes significantly and
even fundamentally, differ, especially in their epistemology.  One
should also be careful as regards Bohr's own versions of
complementarity, of which there is more than one as well.  Here I
shall be specifically concerned with the post-EPR version of his
interpretation, by and large finalized by Bohr in his ``Discussion
with Einstein on Epistemological Problems in Atomic Physics'' (1949)
and related later works.\footnote{Niels Bohr, {\it Philosophical Writings
of Niels Bohr}, 3 vols., Ox Bow Press, 1988, vol.  2, pp.  32-66.  This
work will hereafter referred to as {\it PWNB}.} In my view, these works
offer Bohr's most consistent and most worked out exposition of his
argument, in part because this exposition (re)defines what constitutes
{\it individual} physical phenomena that are and, in his view, could only be
rigorously considered by quantum mechanics as, or, again, interpreted
as, complementarity.  All relevant experimental data must be seen in
accordance with this definition.  There are also different
{\it interpretations} of complementarity itself, as Bohr's interpretation of
quantum mechanics, whichever version of complementarity one considers. 
This is of course also true with respect to other interpretations of
quantum mechanics, since whenever we consider such interpretations we
also interpret them.  Accordingly, the present argument represents
only an interpretation of Bohr's interpretation of quantum mechanics,
even leaving aside those points of my argument whose attribution to
Bohr's view is uncertain, or of course those that expressly depart
from Bohr.  I shall indicate these points as I proceed.  

Interpreted as complementarity, quantum mechanics is seen as a theory
that a) employs a mathematically rigorous formalism; and b) predicts,
in statistical terms, the outcome of all relevant experiments within
the scope of phenomena it considers.  In addition, there is no
conflict between quantum mechanics and other experimental data of
physics, in particular (this becomes crucial) relativity.  Quantum
mechanics is a local theory.  

In this sense, quantum mechanics is as complete as classical physics. 
Complementarity, however, makes quantum mechanics {\it epistemologically}
different from classical physics, insofar as, in this interpretation,
quantum mechanics only {\it predicts} the outcome of relevant experiments
but {\it does not describe}, even in principle and even as an idealization,
the physical behavior of the ultimate objects that it investigates,
quantum objects, and that are responsible for the appearance of the
data in question.  (Classical physics, specifically classical
mechanics, expressly does both, at least in idealized cases and, one
might add, in most interpretations of it, including by Bohr.  For one
could in principle interpret classical physics differently.)  Indeed
complementarity sees such a description as, in principle, impossible. 
This view establishes the difference between Bohr's and more
traditionally positivist approaches, according to which, roughly
speaking, one merely need not be concerned with providing such a
description.  This difference is epistemologically crucial, even
though it may not be significant from the point of view of handling
physics.  Classical statistical mechanics also does not describe the
behavior of the individual constituents, such as molecules, of the
multiplicities it considers, but these constituents are assumed to
behave classically, specifically, according to the laws and pictures
of classical mechanics, which assumption is indeed crucial for the
probability laws and counting procedures of classical statistical
mechanics.  This assumption, however, is not possible in
complementarity in any circumstances, either in the case of individual
objects or quantum collectivities, which also restrict quantum
mechanics to statistical predictions concerning such behavior, or
again, more accurately, to the impact of such behavior upon measuring
instruments.  Quantum mechanics can make some exact predictions,
constrained by uncertainty relations, for example, in the EPR
situations, but by virtue of the limitations imposed by uncertainty
relations, such predictions are, from the perspective of
complementarity, never sufficient to allow one to assume a
classical-like behavior of quantum objects.  Nor, in this view, can
one rigorously speak of quantum objects themselves either as particles
or as waves, or in terms of even partial properties of particles or
waves, or indeed in terms of any conceivable physical objects or their
attributes.  The concept itself of an individual event or phenomenon
needs to be redefined in terms of observable effects of a given
interaction between quantum objects and the measuring instruments
involved, an interaction that is accordingly, irreducible, in contrast
to what obtains in classical physics.  This redefinition also leads to
a particular view of uncertainty relations, which, as I shall also
explain below, now rigorously apply only to the classical physical
behavior of certain parts of measuring instruments.  

These considerations do not of course imply that ``quantum objects,''
or rather what we infer as such, do not exist.  Quite the contrary; it
is {\it their existence} that, in this view, prevents us from conceiving of
{\it the way they exist} or behave, or, to begin with, makes us infer the
existence of such entities from the experimental data in question.  In
other words, we are compelled to infer this ``inconceivable'' from
certain effects manifest in measuring instruments, such as those of
the double-slit experiments, while the particular character of these
effects, especially considered in their totality, prevents us from
unambiguously speaking of any properties, even single properties, of
quantum objects themselves or of their behavior.  

Indeed, quantum mechanics as complementarity does not physically
describe the behavior of any physical systems, including measuring
instruments, which are, in this interpretation, seen as described by
means of classical physics.  More accurately (there is much
misunderstanding on this point), classical physics describes those,
and only those, parts of measuring instruments where the data in
question is registered.  By contrast, those components of measuring
instruments that interact with quantum objects (it is this interaction
that is responsible for the appearance of the effects in question) are
quantum and, hence, are themselves indescribable, as is the
interaction itself by virtue of its quantum character.  

Of course, if one sees (as Einstein did) a physical description of the
ultimate objects of the theory as a requirement of completeness, then
quantum mechanics as complementarity would be incomplete. 
Accordingly, I call the very possibility, available in and usually
defining classical physics, of such a description ``classical.'' 
(Bohmian theories, for example, would be seen as classical in this
view, even though they should be rigorously distinguished from
Newtonian mechanics.)  In other words, quantum mechanics may be
incomplete by classical epistemological criteria, established largely
on the basis of classical physics or even a particular (causal and
realist) interpretation of classical physics.  This (i.e. classical)
type of interpretation appears much more difficult to attain and is
perhaps (according to complementarity, definitively) impossible in the
case of quantum mechanics.  In any event, short of this requirement,
which is, again, epistemological or philosophical, quantum mechanics
as complementarity is complete within its proper scope, as complete as
classical physics is within its proper scope. 

Is quantum mechanics as complementarity as complete as any theory of
quantum data can possibly be?  I would not make such a strong claim
and I do not think that Bohr, or even Heisenberg, does either,
although their position may, at points, be somewhat stronger than
mine.  (Khrennikov appears to attribute a stronger position to my own
argument as well.)  It is, as I said, true that there have been some
who believed or argued, and some still do, that this type of
interpretation is inevitable, often without actually following the
interpretation itself (be it Bohr's complementarity or other) to its
radical epistemological limits or even giving it a careful
consideration at all. 

It is in view of the considerations just given, especially the
understanding that complementarity is {\it an} interpretation, {\it a
particular interpretation}, of quantum mechanics (the standard
quantum-mechanical formalism cum the data in question), that some of
Khrennikov's arguments concerning complementarity appear puzzling to
me.  For, it follows that, contrary to Khrennikov's apparent view,
complementarity neither prohibits nor aims to prohibit the possibility

a) that quantum mechanics could be interpreted otherwise, as it has
been and continues to be, along all conceivable lines and, if
anything, with increasing profusion; and

b) that other theories (such as those of the Bohmian type) of the same
data could be offered.

Accordingly, Khrennikov's apparent argument to the contrary appears to
me misplaced.  This argument appears to assume that complementarity
prohibits certain epistemological alternatives, while I would argue
that, {\it as an} interpretation of quantum mechanics, complementarity
prohibits certain epistemological (or ontological) features within its
own framework, which other interpretations or theories may allow.  The
assessment of such alternatives themselves is, again, a different
question.

I might add the following point to my argument for complementarity as
{\it an} interpretation, and only an interpretation (one among
possible others), of quantum mechanics.  On the one hand, this
argument may be seen as entailing a weaker claim, again, possibly than
that of Bohr himself or that of Heisenberg, upon complementarity or
quantum mechanics itself as concerns their relations to nature at the
ultimate level of its constitution (including whatever dynamics
behavior of possible constituents might be involved).  Upon this
constitution complementarity makes no claim at all, and, by
definition, it prohibits any such claims as part of its own conceptual
structure.  On the other hand, however, it follows that this {\it
impossibility} of any description or conception concerning this
constitution is itself an {\it idealization} pertaining only to
complementarity and {\it not} to nature itself.  Strictly speaking,
one should, accordingly, put quotations marks around ``nature'' (at
least at that level), or ``quantum,'' ``constituents,'' ``dynamics,''
``ultimate,'' or indeed any term one might possibly use here.  We do
not know whether such terms are ultimately applicable or not to
``nature'' at that level.  They are ultimately not applicable, anymore
than ``particles'' or ``waves,'' in complementarity as an
interpretation of quantum mechanics, and function there only
provisionally or metaphorically.  They may be applicable to a greater
degree in classical-like physical theories or interpretations,
including those of quantum mechanics.  The development of our theories
(for example, quantum field theories) and of their interpretations may
lead to more classical-like views of them and, possibly, of nature
itself.  It may, however, also reveal even great complexities and yet
more radical forms of inaccessibility or inconceivability of
``nature,'' even though it appears difficult, if not impossible, to go
beyond complementarity on that score.  

As concerns complementarity itself and substantive problems it may
have as an interpretation of quantum mechanics, it seems to me that
Khrennikov's paper does not sufficiently take into account several
among the key aspects of this epistemology outlined above.  In
particular, Khrennikov's argument concerning the statistical view of
quantum mechanics (including uncertainty relations) vs.  the view of
it as a description of individual processes or events does appear to
me to pay sufficient attention to one of the most crucial
epistemological aspects of complementarity.  It is a subtle point and
is indeed easy to miss, in part since we tend to think of physical
phenomena as pertaining to physical objects our theories are concerned
with.  Bohr, however, terminologically and substantively distinguishes
observable phenomena, which manifest themselves only in measuring
instruments and which are always classical, and quantum objects, which
are never observable as such or, again, even conceivable in the sense
of any possibility of giving them any conceivable specific
characterization, be it conceptual, physical, or mathematical. 
Einstein, for example, did not adequately consider this aspect of
Bohr's view either.  The question is how one understands or can
possibly understand individual phenomena in dealing with the data of
quantum mechanics.  As I said, Bohr, in part in response to Einstein's
critique, was compelled to redefine the concept of the individual
phenomenon as applied to quantum data.  According to this view,
quantum mechanics as complementarity only deals with individual
phenomena in this sense.  This is why Bohr's post-EPR version of
complementarity is so crucial and why it is perhaps the only fully
developed version of it. 

Now, if by individual processes and events one refers, as Khrennikov
appears to do, to whatever happens at the level of quantum objects
themselves, then, {\it contrary to his claim}, Bohr's complementarity
{\it cannot} be seen, including as regards uncertainty relations, as a
theory of individual processes or events.  This view of it is not
possible for the simple reason that, as explained above, it does not
describe any processes, individual or collective, at the quantum
level, or, as I said, at any level, leaving the description (always
partial) of the behavior of measuring instruments to classical
physics.  Indeed, as I stressed above, complementarity sees such a
description, or any analysis or reference to the properties of quantum
objects or processes, as, in Bohr's words, ``{\it in principle}
excluded'' ({\it PWNB} 2:62).  Complementarity {\it is}, however, a
theory of individual phenomena, and indeed, in terms of physics, a
theory only dealing individual phenomena, in a different sense.  It
provides rigorous predictions concerning occurrences of individual
phenomena in Bohr's sense and only concerning such occurrences.  Each
such phenomenon is defined in terms of a single classical effect, such
as a ``dot'' on the screen in the double-slit experiment, of the
interaction between the irreducibly indescribable quantum objects and
indescribable quantum parts of measuring instruments upon the
classically describable parts of measuring instruments.  This,
crucially, includes a rigorous specification of the conditions of a
given experiment ({\it PWNB} 2:64).  In Bohr's interpretation, the
{\it influence} of these conditions can never be eliminated in
considering the outcomes of quantum-mechanical experiments in the way
it can, at least in principle, be done in classical physics.  Thus, a
dot on the screen in the double-slit experiment would be seen as (a
part of) a different {\it type} of phenomenon depending upon the
possibility of knowing through which slit an ``object'' that left the
dot (as a trace of its collision with the screen) has passed.  (I
underline ``type'' because, ultimately, each phenomenon is different
and indeed unique in this interpretation.)  If seen independently of
the quantum-mechanical context of its appearance, each mark on the
screen in the double-slit experiment would be perceived in the same
way or as the same phenomena in the sense of the philosophical (say,
Husserl's) phenomenology of consciousness.  Such a mark would appear
as the same regardless of the difference in the conditions and, hence,
outcome ('' interference'' or '' no interference'' ) of the
double-slit experiment.  According to Bohr's understanding, however,
each mark is seen as a different {\it individual} phenomenon or as a
part of a different phenomenon depending on and including these
conditions, which are always mutually exclusive in the case, such as
this, of complementary phenomena.  In classical mechanics such
conditions would of course be the same, assuming that we deal with
particles (in the case of waves they would be the same but there would
be no point-like traces and different equations, wave equations, would
apply).  Quantum-mechanical predictions, including numerical
statistical predictions crucially depend on this distinction as well. 
(Mathematically, these predictions may be seen in terms of contextual
probability in Khrennikov's sense, although the physical content would
be seen differently from the point of view of complementarity.)  Thus,
in the double-slit experiment, rather than dealing with two phenomena,
each defined by a different multiplicity of spots on the screen, we
deal with two distinct multiplicities of individual phenomena, defined
by each spot.  Each of the latter is indivisible---two sets of
phenomenal atoms or atomic phenomena in Bohr's nonclassical
sense---depending on two different sets of conditions of the
experiment.  One of these sets of conditions will lead to the
emergence of the interference pattern, ``built up by the accumulation
of a large number of {\it individual} processes, each giving rise to a
small spot on the photographic plate, and the distribution of these
spots [following] a simple law derivable from the wave analysis''
({\it PWNB} 2: 45-46).  Each single spot, however, must be, again,
seen as a different individual phenomenon, which depends on the
conditions in which the event occurs.  Two different overall patterns,
'' interference'' and ``no-interference,'' pertain, thus, to two (very
large) sets of different individual phenomena.  Far from being a
matter of convenience, this distinction between two multiple-spot
phenomena and two multiplicities of spot-phenomena is essential for
Bohr's meaning and the consistency of his argumentation.  First, no
paradoxical properties, such as simultaneous possession of
contradictory wave-like and particle-like attributes on the part of
quantum objects themselves, are involved.  Secondly, and perhaps most
crucially, in our analysis we can never mix considerations that belong
to complementary experimental set-ups in analyzing a given
experimental outcome, even when dealing with a single spot on the
screen, as we could, in principle, do in the case of classical
physics.  This is not an uncommon error (at least as Bohr's
interpretation and arguments are concerned), including in some of
Einstein's arguments, which could indeed lead to the appearance of
paradoxes, on which point I shall comment in the next section in the
context of the EPR argument and counterfactual logic.  The latter,
however, disappear once this rule of complementarity as mutual
exclusivity of such considerations is followed.  Throughout his
arguments with Einstein, Bohr stresses in such situations, which are
invoked in most of Einstein's arguments, including of the EPR type, ''
we must realize that \ldots we are not dealing with a {\it single}
specified experimental arrangement, but are referring to {\it two}
different, mutually exclusive arrangements'' ({\it PWNB} 2:57).  These
considerations clearly pertain to the question of counterfactual
reasoning in quantum mechanics.  They are also crucial in the context
of contextual probabilities and, especially, their physical
interpretation, and are, I would argue, not sufficiently addressed in
Khrennikov's paper.

Any further analysis of such phenomena is, again, ``{\it in principle}
excluded.''  Accordingly, complementarity makes the
unknowable---something that cannot in principle be known---part, indeed an
irreducible part, of the knowledge that it provides, since this
unknowable has fundamental effects upon what we can know.  These
effects are the data of quantum physics, according to complementarity. 
Given the particular character of the totality of such effects (e.g.
in the double-slit or the EPR experiment), most such predictions, Bohr
argues, can only be statistical.  As Bohr says, however, ``It is most
important to realize that the recourse to probability laws under such
circumstances is essentially different in aim from the familiar
application of statistical considerations as a practical means of
accounting for the properties of mechanical systems of great
structural complexity [as, for example, in classical statistical
physics].  In fact, in quantum physics, we are presented not with
intricacies of this kind, but with the inability of the classical
frame of [physical] concepts to comprise the peculiar feature of
indivisibility, or `individuality,' characterizing the elementary
processes [defined in terms of phenomena in Bohr's sense, rather than
in terms of quantum objects themselves and their behavior]'' ({\it PWNB}
2:34).  In other words, it is the irreducibly unknowable nature of
quantum objects and their quantum interaction with the quantum parts
of measuring instruments even in each individual case that are
responsible for the irreducibly statistical character of
quantum-mechanical predictions.  By the same token, why a particular
statistical counting works in quantum mechanics is in turn beyond the
possible purview of the theory (interpreted as complementarity), in
contrast to classical statistical mechanics, or elsewhere in classical
physics where statistical predictions apply (as in predicting the
probability in throwing dice.  There the statistical behavior and
particular ways of counting are explained from the classical,
Newtonian, causal behavior of each individual constituent of a given
multiplicity, and their interactions, although the large number and
the structural complexity of the components (be they objects, forces,
factors, or whatever) make it impossible to describe each such
constituent.  In quantum mechanics, as complementarity, this type of
description is never possible even when the maximal information
concerning any individual situation is available, and, again, no
description of any kind concerning quantum objects themselves or their
quantum interactions is ever possible.  It is in this (physical) sense
that quantum probabilities are irreducible, the sense, I would argue,
not sufficiently considered in Khrennikov's criticism of Bohr and
complementarity, specifically as concerns the question of probability. 

Some exact predictions are possible in quantum mechanics.  They are,
however, never sufficient to reconstitute a classical-like situation
even at the level of measuring instruments, whose behavior in these
experiments is, while classically described, subject to an only
partial rather than complete classical treatment.  The reason for this
is that, in view of uncertainty relations, one can only establish at
most half of the (conjugate) variables involved in and necessary for a
complete classical description.  It is worth stressing that
quantum-mechanical statistical predictions involve particular (EPR)
correlations and, hence, a certain form of order not found in
classical physics.  Accordingly, as indeed became apparent already in
Planck's Law, the statistical counting in all quantum theories,
including in quantum statistical theories, is not the same as in
classical statistical physics, even though mathematically one deals
with probabilities in both cases.  (Whether quantum probability is
Kolmogorovian has been a matter of some debate, however.  I leave
aside the question of the relationships between Kolmogorovian
probability and contextual probability in Khrennikov's sense.)  While
each individual event, such as registering a ``dot'' or a ``click''
(it actually has a complex structure), is a random event, the
collective accumulation (a temporal process!)  exhibits a certain
pattern, an '' interference pattern'' (using this term with caution,
since there are no physical waves involved), in a double-slit
experiment.  In this sense, again contrary to Khrennikov's argument,
complementarity is a statistical interpretation.  It is an irreducibly
statistical theory of both individual and collective (rather than only
collective) physical phenomena. 

The uncertainty relations fit nicely into this scheme.  Any
measurement could only involve a determination and indeed, according
to Bohr, a {\it definition} of one of the two conjugate variables for the
relevant part of the measuring instrument involved (once again,
nothing else is subject to measurement or prediction in
complementarity), while Heisenberg's formula would require statistical
confirmation and may be also derived statistically.  Accordingly, it
does not appear to me that there is any logical inconsistency here, as
Khrennikov appears to suggest.  Let me note that to some degree,
uncertainty relations are a remnant of classical mechanics. 
Technically, one can see quantum mechanics or, indeed even more so,
quantum field theory, as a theory of statistical distributions of
certain individual events (dots, spots, traces, clicks, etc.), a
theory predicting such distributions, without any appeal whatsoever to
uncertainty relations or any mechanical variables of quantum objects. 

I do see in Khrennikov's paper an argument that the epistemology of
complementarity is unsatisfying (basically Einstein's point) or at
least that one does not have to accept it and that alternatives could
be offered, or, in view of this dissatisfaction, should be (also
Einstein's point) offered.  Indeed, Khrennikov's main desiderata are
very much those of Einstein.  Quantum mechanics was acceptable to
Einstein as a statistical theory, while he hoped that a
classical-like, realist and preferably causal, alternative describing
the individual behavior of individual quantum objects could eventually
be found.  Einstein's analysis of the situation itself was different
from Khrennikov's in that Einstein did not see how quantum mechanics
in its standard version could account for individual events and
processes at the quantum level, that is, in terms of independent
properties of quantum objects and processes, at least without
violating locality/relativity.  (I put aside for the moment the role
of the EPR correlations in shaping his view, which I have considered
in the works cited above, and I, again, permit myself to refer to
them.)  Einstein saw quantum mechanics, including in Bohr's
interpretation, as irreducibly statistical in the sense of dealing
only with ensembles (of quantum entities).  Bohr tried to counter,
along the lines here sketched, that it was, in his interpretation,
also a theory of individual classical phenomena, that made any account
of such phenomena (i.e. predictions concerning them, since nothing
else is possible in Bohr's interpretation) unavoidably statistical. 
He was not very successful in this argument (Einstein accepted only
some of Bohr's points).  This is hardly surprising since Einstein was,
at best, reluctant to see any theory not dealing with its ultimate
objects, in this case quantum objects, as ultimately acceptable.  He
found Bohr's view `'so very contrary to his scientific instincts,''
the statement that Bohr does not fail to cite ({\it PWNB} 2:61).  Thus
interpreted, quantum mechanics was almost not physics to Einstein
(indeed I am not sure that ``almost'' is necessary here).  In a way,
he was not wrong, since (this is one of my main points in ``Quantum
Atomicity and Quantum Information'' ) Heisenberg's ``new kinematics''
already renounced this type of description and dealt in fact or in
effect with spectra as defined by the effects of the interaction
between quantum objects and measuring instruments along the lines here
outlined.  This ``kinematics'' might have been better called ``quantum
informatics,'' since it was not really kinematics, insofar as it did
not deal with the motion of quantum objects.  Einstein, however, also
accepted that no such alternative was available, either in terms of a
more acceptable interpretation of the quantum-mechanical formalism
(which did not appear to him likely or even possible in view of
locality considerations), or in terms of, at the time, alternative
forms of formalism that would handle the data in question.  He did not
view Bohm's 1952 theory as such an alternative, primarily, I think, in
view of its inherently nonlocal character.  Incidentally,
Schr\"{o}dinger's view of the quantum-mechanical situation was pretty
much the same.  This apparent lack of rigorously developed
alternatives also partially explains why Bohr and Heisenberg were
sometimes less careful than perhaps they should have been as concerns
the fundamentally interpretive character of complementarity, and
sometimes dismissive of counterarguments concerning it, although Bohr
patiently and painstakingly replied to all Einstein's
counterarguments.  

Indeed, no such classical-like (realist and causal) and, this is again
crucial (especially in the context of Bohmian mechanics), also {\it local}
alternative may as yet be available.  This is not to say that it is in
principle impossible, although Bell's and related theorems, such as
the Kochen-Specker theorem, appear to make it nearly impossible.  I
would not go so far as to argue that there are no rigorously worked
out or viable alternatives to complementarity, especially those
dealing with the standard formalism, which would be difficult to do,
given the profusion of available interpretations, although I am
tempted to think that this is the case.  Indeed, it is not always
clear how much of an alternative some of them really are, once they
are rigorously considered or, to begin with, {\it sufficiently worked out
in terms of their physics}, rather than hypothetical or
semi-hypothetical proposals, which is indeed a crucial point.  I
comment on some among recent interpretations from this perspective in
'' Reading Bohr.''  Most of these interpretations are not causal or
realist in the way Einstein wanted, or in the way some Bohmian
theories are, although they, by and large, also want to dispense with
the role of measuring instruments.  It is, again, another question how
successfully.

Let me say a few words concerning the Bohmian approach(es), which, I
stress, are not quantum mechanics and deploy, {\it in fact or in
effect}, a different mathematical formalism.  I say ``in fact or in
effect'' because certain versions of these theories, such as Bohm's
1952 version, do {\it in fact} use the standard, say,
Schr\"{o}dinger's, formalism, while explicitly assigning trajectories
to particles (in these theories there indeed are particles as well as
waves associated with individual particles), which the standard
quantum mechanics does not do.  This assignment, however, implies the
presence of an additional differential equation not found in the
standard theory.  So, {\it in effect}, the formalism is different in
this case as well.  This equation introduced nonlocality, as well as
causality (to the degree that one can speak of causality given
nonlocality), into the theory, even though the theory, it is worth
keeping in mind, is not classical either (i.e. is not a theory of the
type of Newtonian mechanics and is indeed closer to quantum mechanics
than to classical mechanics).  This---nonlocality---is, in my view,
the primary reason why the Bohmian approach is a minority, a small
minority, view of quantum mechanics, although, as you observe, other
factors play a role as well.  There are arguments, such as those by
Henry Stapp, that the standard quantum mechanics is nonlocal as well. 
These arguments, at least so far, have not been effective or, I would
argue, sustainable.  Accordingly, given the current state of physics,
and specifically the extraordinary success of quantum mechanics and
its extensions in quantum field theory in enabling theoretical
predictions of the outcome of experiments, it does not appear to me
that Bohmian theories are likely to have a greater impact in the
current practice of theoretical physics.  (Their mathematical aspects
are obviously another matter).  I might of course be wrong.  I can
hardly see Bohmian approaches as any less ``romantic'' {\it in
Khrennikov's sense} (I shall explain my emphasis presently) or,
correlatively, any more realist or realistic than Bohr's---quite the
contrary, in many respects it seems to me more romantic and less
realistic, albeit realist.  Bell used similar terms in his criticism
of Bohr et al., on which I commented elsewhere.\footnote{ Arkady
Plotnitsky, {\it Complementarity: Anti-Epistemology after Bohr and
Derrida}, Duke University Press, 1994, pp.  182-85.}

On the other hand, as a scholar of Romanticism (which is my literary
field), I would argue that historically Romanticism, for example that
of H\"{o}lderlin, Kleist, Blake, Shelley and Keats, is indeed
epistemologically closer to Bohr than to most realisms.  But this
Romanticism may be more realistic or even more rigorously realist,
insofar as it is scrupulously attentive to the limits within which the
concept of reality can and must apply, than most professed realisms. 
Both the dead nature and life, or the human mind and culture, may
ultimately be better approached from this Romantic perspective.  One
cannot, however---this perspective also tells us---ever be fully certain.

\section{The Letter of Copenhagen}

In this section I shall comment on several specific points of
Khrennikov's paper, pertinent to the above argument, and most of my
points here indeed follow more or less immediately from that argument. 
The page numbers below refer to those of Khrennikov's paper.  

Andrei Khrennikov (hereafter AK): ``I do not think that understanding
of \ldots [the] {\it contextual structure of physical theories} \ldots [i.e. that
physical theories describe properties of pairs, physical system and
measuring device] really was Bohr's invention'' (p.  2).

Arkady Plotnitsky (hereafter AP): It depends on what one means by
contextual, more specifically, what is the epistemology of a given
form of contextualism.  Indeed, it seems to me that contextualism
needs to be more sharply defined by Khrennikov, especially in physical
terms and in relation to the dependence of the outcome of measurements
on different experimental set-ups (such as those in the double-slit
experiment).  Insofar as contextualism is seen as the argument that
the outcome of experiments, including statistical predictions (for
example, in terms of contextual probabilities), fundamentally depends
on a given experimental set-up, Bohr may well have invented it.  He
would of course question certain physical assumptions implicitly
underlying Khrennikov's argument for the role of contextual
probabilities in quantum mechanics.  

AK: ``It was clear to everybody that physical observables are related
to properties of physical systems as well as measuring devices'' (p. 
2).
 
AP: This statement (which needs to be more sharply formulated in any
event) is true in one, more or less trivial, sense: it has indeed
always been recognized that some measuring devices (even if only human
organs) are necessary to establish and relate (observe, measure, etc.) 
such properties.  On the other hand, in classical physics, this
interaction between physical objects under investigation and measuring
instruments could, in Bohr's words, always be ``neglected or
compensated for,'' at least in principle.  In this sense the relevant
properties of objects, such as, say, the position and the momentum of
a given body, can be considered as independently existing and could,
at least in principle, both be simultaneously determined within a
single experimental arrangement, although in practice different
arrangements are often used.  By the same token, while classical
physical theories must indeed rely on measuring instruments to confirm
their descriptions and predictions, they describe, at least in
principle, independent properties of physical systems under
investigation, and predict the values of these properties.  As I have
explained above, this is not the case in quantum mechanics; and, this
is, again, the crux of Bohr's argument, where it is, at least in his
interpretation, impossible to refer unambiguously to such properties. 
It is, in principle, impossible to dissociate quantum objects from
their interaction with the measuring instruments involved and isolate
them in the way it is possible, at least in principle, in classical
physics.  In Bohr's interpretation, there are neither waves nor
particles, nor indeed anything specifiable at the quantum level.  One
deals only with particular effects of the interactions between
``quantum objects'' and measuring instruments upon those instruments,
if one can still, in all rigor, even use such terms as ``quantum'' and
``objects,'' although one can rigorously speak of measuring
instruments, that is, their classically describable parts.  The
fundamental dependence of the outcomes of predictions on specific
experimental set-ups is an immediate consequence.  This,
epistemologically radical or, as I like to call it, nonclassical,
contextualism may well be Bohr's (and Heisenberg's) invention.

AK: ``The main invention of N. Bohr was not contextuality, but
{\it complementarity}.  Bohr's greatest contribution was the recognition of
the fact that there exist complementary experimental arrangements and
hence complementary, incompatible, pairs \ldots .  I think nobody can be
against the recognition of such a possibility.  Why not?  Why must all
contexts, complexes of physical conditions, be coexisting? 
Contextuality and complementarity are two well understandable
principles (not only of quantum physics, but physics in general)'' (p. 
2). 

AP: This statement does not seem to me to be altogether accurate. 
First of all, it, again, depends on what kind of contextuality one has
in mind.  As I said above, both contextuality (in the sense of the
irreducible dependence of the outcome of an experiment on the
experimental set-up and, as a result, the rigorous impossibility of
ascribing anything to quantum objects or considering them apart from
their interactions with measuring instruments) and complementarity (in
Bohr's rigorous sense of the irreducible mutual exclusivity of
measuring procedures) are, according to Bohr's interpretation,
irreducible only in quantum, but not in classical, physics.  In
particular, as I said, in classical physics it is, at least in
principle, possible to measure both the position and the momentum of a
given body within the same experimental arrangement, although in
practice different arrangements are often used.  This is never
possible in quantum mechanics in view of the uncertainty relations. 
It may be worth stating this point in more rigorously Bohrian terms,
defined by the fact since no properties, such as those of position and
momentum, could be attributed to quantum objects, not even each such
property by itself.  No quantum-mechanical measurement allows one both
to fix both a space-time reference frame for any position
determination and to measure a change of the momentum of the relevant
parts of measuring instruments, in the way it could be done when we
measure a classical object in classical physics.  In quantum
measurement (in this interpretation) we can do either one or the
other, but never both together.  Both this radical type of
``contextuality'' (this is of course not Bohr's term, perhaps because
in itself it does not convey the epistemologically radical nature of
his idea) and complementarity may be seen as Bohr's conceptions.

AK: ``The real problem was that N. Bohr as well as W. Heisenberg (but
merely further generations of their adherents) did not pay attention
that quantum complementarity is the experimental fact concerning pairs

\begin{center}
$ \pi $ = (elementary particle, macroscopic measuring device)
\end{center}

and not elementary particles by their sel[ves].  It is
a pity that the greatest promoters of contextualism forgot about
contextual basis of complementarity'' (p.  2).

AP: This is manifestly incorrect, since, as was shown in the preceding
discussion, Bohr's complementarity only concerns experimental
arrangements in their interaction with ``quantum objects.''  (It can
be shown that Khrennikov's statement is not true about Heisenberg
either.) 

AK: ``\ldots complementarity of contexts in quantum physics does not imply
complementarity of corresponding objective properties (of elementary
particles) contributing into such observables'' (p.  2). 

AP: Bohr, as I explained above, never said that it would or, again,
could, given that no such properties could be ascribed to quantum
objects. 

AK: ``In particular, contextual complementarity does not imply that
elementary particles do not have objective properties at all.  In
particular, there are no reasons to suppose that it is impossible to
provide a kind of hidden variable, HV, description ({\it ontic description}
\ldots) for these objective properties'' (p.  2). 

AP: As I said, for Bohr complementarity does in fact imply precisely
this impossibility of speaking of independent objective properties of
quantum objects, as Khrennikov indeed acknowledges later in the
article.  Of course, as I also said, this does not prevent one from
making this type of presupposition, trying to pursue it, and connect
it with, in Khrennikov's terms, ``contextual complementarity.''  This
is, for example, Bohm's program.  The question is how successful one
is in this pursuit.  Bohm's theory may indeed be seen as achieving
something of that type, but, again, at the expense of introducing
nonlocality as a mathematical consequence of the theory.  Hence, also
the significance of Bell's and related theorems in assessing such
projects.  And, as I said, it is not clear to me how {\it p}-adic
hidden variable theories fare in this respect, or how they actually
relate to experimental results, to begin with.  Perhaps, they can
reproduce quantum-mechanics predictions while avoiding nonlocality. 
Can they?  Even if they can, it would still leave other problems as
concerns relating this type of mathematics to anything physical.  As
Khrennikov acknowledges, this remains a hypothetical proposal: ``the
development of alternative (nonreal, noncontinuous) classical models,
e.g. {\it p}-adic \ldots, {\it might} play important role in
clarification of foundations of quantum theory'' (p.  3; emphasis
added).  They might, but then they also might not.

AK: ``On the other hand, an adherent of N. Bohr [Plotnitsky] would
argue that `Such a separation and, hence, the description of
(properties of) quantum objects and processes themselves (as opposed
to certain effects of their interaction with measuring instruments
upon latter) are impossible in Bohr's interpretation,' '' (p.  3).

AP: Obviously, I agree, since this is indeed what I say; and I am
grateful to Khrennikov for his choice of quotation here.  I only
direct one's attention to the significance, a fundamental
significance, of my qualification ``in Bohr's interpretation,'' which
must be viewed along the lines outlined above and which does not
appear to me to be sufficiently taken into account by Khrennikov.

AK: ``I think that the origin of such an interpretation of
complementarity by N. Bohr was the individual interpretation of
Heisenberg's uncertainty principle [uncertainty relations?]:
$ \Delta q \Delta p \geq h/2 $.''

AP: As I said, Khrennikov's discussion of uncertainty relations
appears to me to misconceive Bohr's or (admittedly, this would require
a separate discussion) Heisenberg's view, as explained above and in
more detail in my works cited earlier.  In particular, Khrennokov
says: ``Unfortunately, Heisenberg's uncertainty relation was
interpreted as the relation for an individual elementary particle''
(p.  3).  First of all, one might ask: By whom, specifically?  It was
by some, to be sure, but certainly not by Bohr, at least not in this
sense, but only in the sense of individual phenomena, as explained
above, that is, in relation to certain parts of measuring instruments
in two possible but always mutually exclusive experimental
arrangements.  I leave aside Khrennikov's historical commentary, which
is far too sketchy and incomplete, and is not really germane to the
argument.

AK: ``The main problem was mixing by W. Heisenberg of {\it individual} and
{\it statistical} uncertainty.  For example, in his famous book [{\it The
Physical Principles of Quantum Theory}] he discussed the uncertainty
principle as a relation for an individual system, but derived this
principle by using statistical methods!''  (p.  3).

AP: This sounds to me confused, at least as stated, including as
concerns the difference between the uncertainty principle and
uncertainty relations, and, in any event, this point is not
sufficiently explained by Khrennikov.  As I said, one can, and in
quantum mechanics perhaps must, rigorously combine individual and
statistical considerations.  So it is not clear why this is a problem
or what the problem in fact is.  This needs to be explained further
as, in my view, does most of Khrennikov's commentary on uncertainty
relations.

AK: ``In any case the absence of continuous classical model for motion
of electron in Bohr's atom does not imply impossibility to create
other, noncontinuous, classical (causal deterministic) models'' (p. 
4).

AP: Yes, but, again, the question is how effective such models are,
and indeed whether we really do have them or whether we can seem them
as classical, and in what sense.  Indeed, it is quite clear that
``mixing'' or, I would say, {\it conjoining} the classical and the
continuous is not as misconceived as Khrennikov appears to think (pp. 
4-5).

AK: ``Moreover, considerations of W. Heisenberg \ldots [in his
original matrix mechanics paper] did not [even] imply [the]
impossibility to create [a] continuous classical model---as it was
claimed by W. Heisenberg and then by N. Bohr.  The story is much
simpler: first Bohr tried to create such a thing, but could not; then
Heisenberg, with the same result.  After this it was claimed that such
a model did not exist.  And what is the most interesting: not only for
Bohr's model of atom (well it might be), but for any other model\ldots  I
cannot understand this kind of `quantum logic' '' (p.  4).

AP: Again, claimed by whom, specifically?  The nature of Bohr's and
Heisenberg's claims was, it seems to me, more complicated.  But in any
event, my argument above concerning the interpretive nature of
complementarity obviously leaves space for a search for alternatives,
even if, again, my view may stress the interpretive nature of
complementarity more than others (possibly including Bohr and
Heisenberg).

AK: ``So Bohr's complementarity was a kind of {\it individual
complementarity}.  Complementary features were regarded [as belonging?] 
to individual physical systems.  It is a pity that contextualists N.
Bohr and W. Heisenberg related the uncertainty relation[s] not to some
special class of measurement procedures of the position and momentum
described by quantum formalism, but to the position and momentum of an
individual elementary particle'' (p.  5).

AP: As shown by the above argumentation, this is manifestly incorrect
as stated, and appears to flatly disregard both Heisenberg's and
Bohr's arguments, extending from Heisenberg's new kinematics to Bohr's
post-EPR arguments for complementarity.

AK: ``This imply the prejudice that the position and momentum even in
principle could not be determined simultaneously and, moreover, that
it is even in principle impossible to assign such a physical property,
e.g. position or momentum, to e.g. electron: `electron does not have
trajectory' '' (p.  5).

AP: This is, in my view, not simply a prejudice (a view in part
resulting perhaps from the preceding incorrect statement of
Khrennikov's paper, just cited), but, at least, {\it an} interpretation, and
alternatives are not easy to offer.  Bohmian theories, which have
trajectories, are nonlocal, and indeed they are nonlocal because they
have trajectories.  And then, can one speak {\it physically} of {\it
p}-adic, and hence discontinuous, {\it trajectories}?  This may not be
impossible, but it is not easy, assuming, again, that such theories
could be properly developed, including as classical.

AK: ``In fact, the only possible conscious [?]  interpretation of
Heisenberg's uncertainty principle is the statistical contextual
interpretation \ldots.  It is impossible to prepare such an ensemble
of elementary particles that dispersions of both position and momentum
observables would be arbitrary small.  Everybody would agree that only
this statement can be verified experimentally.  Contextualism has to
be {\it statistical contextualism} and, consequently, complementarity has to
be {\it statistical contextual complementarity}.  Such contextualism and
complementarity do not contradict to the possibility of [a] finer
description of reality than given by quantum theory'' (pp.  5-6).

AP: It is true that they do not contradict this possibility.  But,
again, a possibility is not an actuality.  It is Bohr's {\it interpretation}
of uncertainty relations that at any given point only one of the two
conjugate variables involved (pertaining to certain parts of the
measuring instruments used) could be {\it defined} rather than only measured
with full precision within the capacity of our measuring instruments. 
This interpretation arises from his analysis of quantum measurement,
including in the EPR case.

AK: ``In particular, our contextual probabilistic investigations
demonstrated that contextual complementarity, wave-particle dualism,
is not rigidly coupled to microworld.  Thus we can, in principle,
perform experiments with macro systems that would demonstrate
`wave-particle duality,' but not of macro objects, but contexts'' (p. 
7).

AP: Although one might want to see an example of such an experiment to
be able to assess it, I merely observe here that there is no such
duality in Bohr's interpretation, and especially there are no
(continuous) waves, as must be clear already in the double-slit
experiment.  I must add that this statement needs more lucidity in any
event.

AK: ``We have two theoretical descriptions of this experiment: 1)
quantum-like statistical description; 2) Newtonian classical
description.  Both theories give the same statistical distribution of
spots on the registration screen.  Quantum-like theory operates with
complex waves of probability; there is uncertainty, Heisenberg-like,
relation for position and momentum.  Of course, this relation is the
statistical one.  Suppose now that some observer could not provide the
verification of Newtonian description, e.g. such an observer is a
star-size observer and its measuring device produce nonnegligible
perturbations of our macroscopic charged balls.  Such an observer
might speculate on [the] impossibility to find objective phase-space
description and even about waves features of macroscopic balls'' (p. 
7).

AP: To repeat (this qualifications seems to be necessary at any point
of Khrennikov's paper), it is one thing to presuppose the possibility
of a classical-like account and it is quite another to actually
develop one.  In addition, Khrennikov's argument here seems to me
either incorrect or insufficiently developed.  I do not think that
``both theories [quantum and classical] give the {\it same} statistical
distribution of spots on the registration screen.''  For, if one
could, even in principle, count electrons, say, those passing through
the slits in the double-slit experiments as Newtonian balls---as we
indeed can, but only in one among the two possible complementary
arrangements---there would not be an interference-like pattern of dots
resulting from collisions between electrons and the screen.  Thus,
complementarity---the necessity of two different, mutually exclusive,
arrangements for a comprehensive account of the situation we encounter
in quantum physics---changes everything here or, rather, precisely
reflects the difference, possibly irreducible, between the classical
and the quantum.  A Newtonian theory, at least any available Newtonian
theory, does not appear to me to be able to account for this
situation.  Bohmian theories, which do account for it, but nonlocally,
are {\it not} Newtonian.

There may also be an element of terminological and possible
substantive confusion here, insofar as the ``classical''
configurations that Khrennikov refers to are in fact
computer-generated models developed in discrete rather than continuous
time and their status as classical models (if they indeed could be
seen as classical) requires further explication.  Let me, however,
offer an argument, via Anthony J. Leggett's elegant exposition,
describing a different but equivalent experiment (in neutron
interferometry), in which instead of slits we consider the initial
state A, two intermediate states B and C, and then a final state E.
(The latter is analogous to the state of a ``particle'' at the point
of its interaction with the screen in the double-slit experiment.) 
First, we arrange to block the path via state C, but leave the path
via state B open.  (In this case we do not attempt to install any
additional devices to check directly whether the object has in fact
passed through state B.) In a large number (say, again, a million) of
trials we record the number of particles reaching state E. Then we
repeat the same number of runs of the experiment, this time blocking
the path via B, and leaving the path via C open.  Finally we repeat
the experiment again with the same number of runs, now with both paths
open.  In Leggett's words, ``the striking feature of the
experimentally observed results is, of course, summarized in the
statement that \ldots the number reaching E via `either B or C'
appears to be unequal to the sum of the numbers reaching E `via B' or
`via C'.''\footnote{ Anthony J. Leggett, ``Experimental Approaches to
the Quantum Measurement Paradox,'' {\it Foundations of Physics}, 18, 9
(1988): 940-41.} The probabilities of the outcomes of individual
experiments will be affected accordingly.  (This is of course also the
situation that leads Khrennikov to his use of contextual probability.) 
The situation is equivalent to the emergence of the interference
pattern when both slits are open in the double-slit experiment.  In
particular, in the absence of counters, or in any situation in which
the interference pattern is found, one cannot assign probabilities to
the two alternative ``histories'' of a ``particle'' passing through
either B or C on its way to the screen.  If we do, the above
probability sum law (based in adding the so-called amplitudes, related
to the wave-function, to which one applies Schr\"{o}dinger's equation,
rather than, as in the classical case, probabilities themselves) would
not be obeyed and the conflict with the interference pattern will
inevitably emerge, as Bohr stressed on many occasions.  This is also
why the ways of counting probabilities are so different in classical
and quantum physics, as Planck discovered.  One may also put it as
follows.  We must take into account the possibility of a particle
passing through both states B and C (and through both slits in the
double-slit experiments), when both are open to it, in calculating the
probabilities of the outcomes of such experiments.  We cannot,
however, at least in Bohr's interpretation, assume either that such an
event in space and time physically occurs for any single particle,
anymore that we can assume that one can walk into a building
simultaneously through two doors, when these doors are sufficiently
far apart.  The inherently quantum-mechanical nature of the EPR
correlations may be linked to similar considerations as well.  These
considerations lead to considerable complications in Bohmian theories,
which indeed need to have both particles (having trajectories) and
waves, associated with individual particles, to reproduce the
predictions of the standard quantum mechanics, but, again, as a result
introduce nonlocality into the theory.  It seems to me that much of
Khrennikov's argument does not pay sufficient attention to these
considerations or at least does not sufficiently address them.

It would be difficult to speculate how a possible ``intelligent''
observer with perceptual and reasoning capacities analogous to our own
but of the size of a star would see the physical world in our
universe, for example, an object of the size of a billiard ball.  A
more interesting question may be whether there could be ``observers''
(it is not inconceivable that they actually need to be very ``large''
rather than small) in our universe who would in fact `'see'' anything
on the quantum scale, that is, the scale defined by Planck's constant
$ h $.  I would think that, if they could, the picture would be quite
fantastic by our standards.  Perhaps, however, an irreducibly
classical perception, such as our own (defined by the particular
constitution of our bodies) is necessary in order to {\it observe} anything
to all.  My intuitive guess (which could of course be wrong) is that,
given the ultimate physical nature of our universe, different possible
``intelligent'' beings in it might see different ``classical-like''
worlds, but could never see ``the quantum world,'' if, again, such
terms as ``quantum'' and ``world'' could apply.

AK: ``Finally, we remark that the possibility of $ H3 $-description
implies that `quantum randomness' does not differ essentially from
`classical randomness.' Of course, this contradict[s] to[?]  orthodox
quantum views to randomness as {\it fundamental or irreducible
randomness}.  Unfortunately, I could not understand the latter ideas. 
Instead of fundamental irreducible quantum randomness, I prefer to
consider well understandable theory of context (complex of
experimental physical conditions) depending probabilities'' (p.  7). 

AP: This, at the very least, needs to be further explicated.  It seems
to me that Khrennikov's argument is in essence of the Bohmian type,
insofar as the statistical nature of quantum mechanics may only
reflect our partial knowledge of a more classical-like configuration. 
This, again, may or may not be true, but so far there have been no
effective arguments (at least for me and most physicists), in part in
view of the nonlocality of Bohmian theories, that such is the case. 
As I said above, one would still need to account for the difference in
counting, reflected already in Planck's law.  In Bohr's
interpretation, in any event, contextuality entails irreducible
randomness in the physical sense explained above, and not really
addressed by Khrennikov.

In his section, ``Citation With Comments,'' Khrennikov presents `'some
citations on orthodox quantum theory and our contextual statistical
realist comments.  We use, in particular, collections of Bohr's views
presented in papers of H. Folse and A. Plotnitsky'' (pp.  7-8).  I
think that part of the problem of this section is that the statements
Khrennikov discusses are taken from different works, which are not
always linked by Bohr and reflected very different stages of Bohr's
thinking.  The purpose of this section of the article is unclear to
me.  I shall, however, comment on it in order to clear up some
misunderstandings (some of them are not uncommon) that appear to me to
affect Khrennikov's commentary.

AK: (S1) ``In contrast to ordinary mechanics, the new quantum
mechanics does not deal with a space-time description of the motion of
atomic particles.  It operates with manifolds of quantities which
replace the harmonic oscillating components of the motion and
symbolize the possibilities of transition between stationary states in
conformity with the correspondence principle,'' N. Bohr.

This is simply the recognition of the restrictiveness of the domain of
applications of quantum theory.  I would like to interpret this as the
recognition of incompleteness of quantum theory.  However, it was not
so for N. Bohr (p.8).

AP: Bohr specifically describes Heisenberg's ``new kinematics'' here,
which for the first time enabled correct predictions of the key
experiments concerning electrons through a fully developed
mathematical scheme, as opposed to the patchwork of the old atomic
theory of Planck, Einstein, Bohr, and Sommerfeld.

AK: (S2) `` \ldots  the quantum postulate implies that any observation of
atomic phenomena will involve an interaction with the agency of
observation not be neglected.  Accordingly, an independent reality in
the ordinary physical sense can neither be ascribed to the phenomena
nor to the agencies of observation,'' N. Bohr. 

The first part of this citation is the manifestation of contextuality. 
However, I cannot understand what kind of logic N. Bohr used to
proceed to the second part.  The second part can be interpreted as the
declaration of the impossibility of objective, ontic description of
reality (p.8).

AP: Bohr's explains this more rigorously later in the same article,
and better in his later works, along the lines explained above; this
is cited from his Como lecture, where these ideas are as yet somewhat
tentative.  But even the Como lecture provides further explanation. 
It is very difficult to consider this and indeed most of Bohr's
statements apart from his overall argument.

AK: (S3) ``\ldots to reserve the word phenomenon for comprehension of
effects observed under given experimental conditions \ldots These
conditions, which include the account of the properties and
manipulation of all measuring instruments essentially concerned,
constitute in fact the only basis for the definition of the concepts
by which the phenomenon is described,'' N. Bohr.

I would agree if the last sentence would be continued as ``is
described in quantum formalism'' (p.  8).

AP: Bohr only deals with quantum mechanics here.  Khrennikov appears
to miss the {\it context} of Bohr's statement.  There was no other formalism
at the time and it was this formalism that was used by EPR, to whom
this argument replies.  Indeed, rigorously speaking, such phenomena
are, as I said, only predicted by using the quantum-mechanical
formalism, and are physically described classically.

AK: (S4) ``\ldots by the very nature of the situation prevented from
differentiating sharply between an independent behaviour of atomic
objects and their interaction with the means of observation
indispensable for the definition of the phenomena,'' N. Bohr.  I would
agree if the last sentence would be continued as ``of the phenomena
described by quantum formalism'' (p.  8).

AP: See my comment above, specifically my discussion of
complementarity as only {\it an} interpretation of quantum mechanics.  And,
again, these phenomena are, in Bohr's view, definitely {\it not described}
by quantum formalism, which, in this view, describes nothing, nothing
physical, at all.

AK: ``Bohmian mechanics \ldots ---well, it has its disadvantages, but
merely, mathematical'' (p.  9).

AP: I would take issue with this assessment.  The main (although
indeed no the only) disadvantage of Bohmian theories is, in my view,
their nonlocality, which is a physical feature, albeit a mathematical
consequence, of the theory.

AK: ``Finally, Bell's inequality arguments were interpreted as they
should be interpreted in the orthodox quantum framework, despite very
strong counter-arguments.  If all these counter-arguments be taken
into account, Bell's inequality activity would look very strange, as a
kind of mystification.''  AP: This is extremely unclear.  The subject
requires a great deal more explication in any event.  It is indeed
crucial to the problematic in question, and to the ongoing debate
concerning quantum mechanics and the spirit and the letter of
Copenhagen.

\vspace{0.2cm}
I would like to thank Andrei Khrennikov for this opportunity to
discuss the spirit and the letter of Copenhagen, for other productive
exchanges on quantum theory and related subjects, and for organizing
two recent conferences, {\it Quantum Theory: Reconsideration of Foundations}
(2001) and {\it Foundations of Probability in Physics} (2002), at V\"{a}xj\"{o}
University and for inviting me to participate in them.  I am grateful
to other participants of these conferences for many helpful
discussions.

    \end{document}